\documentstyle{cupconf}

\ifoldfss
\else
  \ifnfssone
    \newmathalphabet{\mathit}
      \addtoversion{normal}{\mathit}{cmr}{m}{it}
      \addtoversion{bold}{\mathit}{cmr}{bx}{it}
    \newmathalphabet{\mathcal}
      \addtoversion{normal}{\mathcal}{cmsy}{m}{n}
    \else
    \ifnfsstwo
    \fi
  \fi
\fi

\begin{document}
\ifnfssone
\else
  \ifnfsstwo
  \else
    \ifoldfss
      \let\mathcal\cal
      \let\mathrm\rm
      \let\mathsf\sf
    \fi
  \fi
\fi

\title [Stars in the Galactic Bulge]{Ages And Metallicities For Stars In The Galactic Bulge}

\author[J. A. Frogel]{Jay A. Frogel$1$}
\affiliation{$^1$Department of Astronomy\\ The Ohio State University, 
Columbus Ohio  USA}
\maketitle

\begin{abstract}
     
     Observations of the stellar content of the Milky Way's bulge
helps to understand the stellar content
and evolution of distant galaxies.  In this brief overview I will first
highlight some recent work directed towards measuring the history of
star formation and the chemical composition of the central few parsecs
of the Galaxy.  High resolution spectroscopic observations by Ramrez et
al. (1998) of luminous M stars in this region yield a near solar value
for [Fe/H] from direct measurements of iron lines.  Then I will present
some results from an ongoing program by my colleagues and myself which
has the objective of delineating the star formation and chemical
enrichment histories of the central 100 parsecs of the Galaxy, the
``inner bulge''.   We have found  a small increase in mean [Fe/H] from
Baade's Window to the Galactic Center and  deduce a near solar value
for stars at the center.  For radial distances greater than $1^{\circ}$
we fail to find a measurable population of stars that are significantly
younger than those in Baade's Window.  Within $1^{\circ}$ of the
Galactic Center we find a number of luminous M giants that most likely
are the result of a star formation episode not more than one or two Gyr
ago.

\end{abstract}

\firstsection
\section{Introduction}
     
     The structure and stellar content of the bulge of the Milky
Way are often used as proxies in the study of other galactic
bulges and of elliptical galaxies. However, out to a radius of
about $2^{\circ}$ along the minor axis, and considerably farther
along the major axis, the visual extinction is great enough that
optical observations are difficult to impossible along most lines
of sight. A two degree radius corresponds to 5 to 10" for
galaxies in Virgo. Thus, if we are to use the stellar content of
the inner bulge as a starting point for delineating the global
characteristics of the inner regions of nearby spiral bulges and
spheroidal galaxies, we must turn to near-infrared observations.
The brief review, then, will concentrate on summarizing a some of
the recent studies in the near-IR of the inner bulge of the Milky
Way.  Further, it will specifically address two of the most
important characteristics of the stars:  their ages and
metallicities.

\section{Within A Few Arc Minutes Of The Galactic Center}
     
Krabbe {\it et al.} (1995) have identified more than 20 luminous blue
supergiants and Wolf-Rayet stars in a region not more than a parsec in
radius around the center of the Galaxy.  The inferred masses of some of
these stars approaches 100 M${\odot}$.  From this they conclude that
between 3 and 7 Myr ago there was a burst of star formation in the
central region.  They also identified a small population of cool
luminous AGB stars from which one can conclude that there was
significant star formation activity a few 100 Myr ago as well.

Blum {\it et al.} (1996a) carried out a K-band survey of the central 2
arc minutes of the Galaxy.  They focused on the significant numbers of
luminous ($K_{0} < 6$) cool stars.  These stars are considerably more
luminous than would be expected from a typical old stellar population
such as is found in Baade's Window, for example. Most of these stars
were found by Blum and others to be M stars. With K-band spectra, Blum
{\it et al.} (1996b) were able to distinguish between M supergiants and
AGB stars.  Such a distinction is of importance because of the
implications for the times of star formation.  As first demonstrated
quantitatively by Baldwin {\it et al.} (1973), M-type supergiants can
be easily distinguished from ordinary giants of the same temperature
(or color) via the strengths of the H$_{2}$O and CO absorption bands in
K- band spectra.

Blum {\it et al.} (1996b) found only 3 out of 19 stars to be
supergiants, one of which is the well known IRS 7.  The remainder are
AGB stars. From the spectra and the multi-color photometry they
concluded that there have been multiple epochs of star formation in the
central few parsecs of the Galaxy.  The most recent epoch, less than 10
Myr ago, corresponds with that found by Krabbe {\it et al.} (1995).
Other epochs of star formation identified by Blum {\it et al.} occurred
about 30 Myr, between 100 and 200 Myr, and more than about 400 Myr in
the past.  The majority of stars are associated with the oldest epoch
of star formation.

Abundances for the red luminous stars in the region around the Galactic
Center are being determined by Ramirez {\it et al.} (1998) from a full
spectral synthesis analysis of high resolution K band spectra.  It will
be interesting to compare abundances values for the inner bulge with
their results.  Based on direct measurements of iron lines in 10 stars
they derive a mean [Fe/H] of 0.0 with a dispersion comparable to their
uncertainties, about 0.2 dex. This is only a few tenths of a dex
greater than the mean [Fe/H] determined for Baade's Window K giants
(Sadler {\it et al.} 1996; McWilliam \& Rich 1994). This small increase
in the mean value of [Fe/H] compared with Baade's Window is consistent
with the [Fe/H] gradient in the bulge found by Tiede {\it et al.}
(1995) and Frogel {\it et al.} (1999). The lack of a dispersion in
[Fe/H] contrasts with a dispersion that of more than an order of
magnitude for the K giants in Baade's Window (Sadler {\it et al.} 1996;
McWilliam \& Rich 1994).  It is, however, consistent with the lack of
dispersion found for the M giants in Baade's Window (Frogel \& Whitford
1987; Terndrup {\it et al.} 1991).  The fact that [Fe/H] is near solar at the
Galactic Center with a star formation rate per unit mass that is
considerably in excess of the solar neighborhood value suggests that
the rate of chemical enrichment has been quite different at the two
locations.

\section{The Inner Galactic Bulge}
     
The inner $3^{\circ}$ of the Galactic bulge, interior to
Baade's Window, will be referred to as the inner Galactic bulge.
With the 2.5 meter duPont Telescope at Las Campanas Observatory I
have obtained JHK images of 11 fields within the inner bulge,
three of which are within $1^{\circ}$ of the Galactic Center.  The
two questions to address are:   What is the abundance of the
stars in this region and is there any evidence for a detectable
population of intermediate age or young stars? My collaborators
and I are taking two approaches to the abundance question.  The
first is based on the fact that the giant branch of a metal rich
globular cluster in a {\it K, JK} color magnitude diagram is linear
over 5 magnitudes and has a slope proportional to its optically
determined [Fe/H] (Kuchinski {\it et al.} 1995). Results from work will
be summarized here.  The second approach, which will give a
better answer to the abundance question, is based on the analysis
of K-band spectra of about one dozen M stars in each of 11
fields.  This is a work in progress.

\subsection{Abundances In The Inner Galactic Bulge}
     
     The best "fixed reference point" in any measurement of
abundances within the inner bulge is the determination by
McWilliam \& Rich (1994) of a mean abundance of [Fe/H] = ~–0.2 for
a sample of K giants in Baade's Window based on high resolution
spectroscopy. Sadler {\it et al.}'s (1996) spectroscopy of several
hundred K giants in Baade's Window yielded a similar result.
Both of these analyses measured a spread in [Fe/H] in Baade's
Window of between one and two orders of magnitude. The estimate
of [Fe/H] for the Baade's Window giants based on the near-IR
slope method (Tiede {\it et al.} 1995) differed from previous near-IR
determinations in that they found an [Fe/H] close to the value
based on the optical spectra of K giants.
     
My near-IR survey of inner bulge fields has yielded color- magnitude
diagrams that, except for the fields with the highest extinction, reach
as faint as the horizontal branch.  Thus, with data for the entire red
giant branch above the level of the HB we can apply the technique
developed by Kuchinski {\it et al.} (1995) to determine [Fe/H] from the
slope of the RGB above the HB.  Although the calibration of this
technique is based on observations of globular clusters, the
applicability of this method to stars in the bulge was demonstrated by
Tiede {\it et al.} (1995) in their analysis of stars in Baade's
Window.  This method is reddening independent since it depends only on
a slope measurement.  Based on 7 fields on or close to the minor axis
of the bulge at galactic latitudes between $+0.1^\circ$  and
$-2.8^\circ$ we derive a dependence of $\langle$[Fe/H]$\rangle$ on
latitude for $b$ between $-0.8^\circ$ and $-2.8^\circ$ of $-0.085 \pm
0.033$ dex/degree.    When combined with the data from Tiede {\it et
al.} we find for $-0.8^\circ \leq b \leq -10.3^\circ$ the slope in
$\langle$[Fe/H]$\rangle$ is $-0.064 \pm 0.012$ dex/degree.  An
extrapolation to the Galactic Center predicts [Fe/H] $= +0.034 \pm
0.053$ dex, in close agreement with Ramrez {\it et al.} (1998).  Also in
agreement with Ramrez {\it et al.}, we find no evidence for a dispersion in
[Fe/H].  Details of this work are in Frogel {\it et al.} (1999).

Analysis of the K-band spectra of the brightest M giants in
each of the fields surveyed is nearing completion; the results
appear to be consistent with those based on the RGB slope method,
namely, an [Fe/H] for Baade's Window M giants close to the
McWilliam \& Rich value but with little or no gradient as one goes
into the central region. Also, the spectroscopic data show little
or no dispersion in [Fe/H] within each field.
     
In summary, several independent lines of evidence point to
an [Fe/H] for stars within a few parsecs of the Galactic Center
of close to solar. The gradient in [Fe/H] between Baade's Window
and the Center is small -- not more than a few tenths of a dex.
Exterior to Baade's Window there is a further small decline in
mean [Fe/H] (e.g. Terndrup {\it et al.} 1991, Frogel {\it et al.} 1990;
Minniti {\it et al.} 1995).  It remains to be seen whether this
gradient arises from a change in the mean [Fe/H] of a single
population or a change in the relative mix of two populations,
one relatively metal rich and identifiable with the bulge, the
other relatively metal poor and more closely associated with the
halo.  Support for the latter interpretation is found in the
survey of TiO band strengths in M giants in outer bulge fields by
Terndrup {\it et al.} (1990).  For which they found a bimodal
distribution.  McWilliam \& Rich (1994) proposed an explanation
based on selective elemental enhancements as to why earlier
abundance estimates of bulge M giants seemed to consistently
yield [Fe/H] values in excess of solar. It remains to be
understood why no dispersion is observed in measurements of the M
giant abundances. It also remains to be determined if the
indirect methods used for measuring [Fe/H] are really measuring
iron rather than being sensitive to, for example,  element
enhancements.

\subsection{Stellar Ages In The Inner Galactic Bulge}
     
If a stellar population has an age significantly younger
than 10 Gyr then stars at the top of the AGB will be several
magnitudes brighter than they would in an older population.
After correction for extinction we found that our fields closer
than $1.0^\circ$ to the Galactic Center have significant numbers
of bright, red stars implying the presence of a younger component
to the stellar population, probably with an age of a few Gyr.
This is consistent with Blum{\it et al.}'s work on the inner few arc
minutes of the bulge.  Beyond $1.0^\circ$  from the center there
is no evidence for such luminous stars. Details of this work are
in Frogel {\it et al.} (1999)
    
A second test applied to see if there is evidence for a
young population in the Galactic bulge was a comparison of the
luminosities and periods of bulge long period variables (LPVs)
with those found in globular clusters (Frogel \& Whitelock 1998).
For LPVs of the same age, those with greater [Fe/H] will have
longer periods.  LPVs with longer periods also have higher mean
luminosities.  In the past claims have been made for the presence
of a significant intermediate age population of stars in the
bulge based on the finding of some LPVs with periods in excess of
500-600 days.  It is necessary, however, to have a well defined
sample of stars if one is going to draw conclusions based on the
rare occurrence of one type of star.  The M giants in Baade's
Window are just such a well defined sample (e.g. Frogel \&
Whitford 1987).  Frogel \& Whitelock (1998) demonstrated that with
the exception of a few of the LPVs in Baade's Window with the
longest periods, the distribution in bolometric magnitudes of the
LPVs from the bulge and from globular clusters overlap
completely.  Furthermore, because of the dependence of period and
luminosity on [Fe/H] and the fact that there has been no reliable
survey for LPVs in globulars with [Fe/H] $> -0.25$, the brightest
Baade's Window LPVs could have the same age as the somewhat
fainter ones but come from the higher [Fe/H] population.
     
Finally, observations with the Infrared Astronomical
Satellite (IRAS) at 12$\mu$m were used to estimate the
integrated flux at this wavelength from the Galactic bulge as a
function of galactic latitude along the minor axis (Frogel 1998).
These fluxes were then compared with predictions for the 12 m
bulge surface brightness based on observations of complete
samples of optically identified M giants in minor axis bulge
fields (Frogel \& Whitford 1987; Frogel {\it et al.} 1990).  No evidence
was found for any significant component of 12$\mu$m emission in the
bulge other than that expected from the optically identified M
star sample plus normal, lower luminosity stars. Since these
stars are themselves fully attributable to an old population, the
conclusion from this study was, again, no detectable population
of stars younger than those in Baade's Window, i.e. of an age
comparable to that of globular clusters.

\end{document}